\documentclass[a4paper,12pt]{article}
\title{Sequential optimizing investing strategy with neural networks}
\author{
Ryo Adachi\footnote{Graduate School of Information Science and Technology, 
University of Tokyo, 7-3-1 Hongo, Bunkyo-ku, Tokyo 113-8656, JAPAN}
\footnote{Institute of Industrial Science, University of Tokyo, 4-6-1 Komaba, Meguro-ku, 
Tokyo 153-8505, JAPAN}\\ 
Akimichi Takemura\footnotemark[1]
}
\date{February 2010}
\usepackage{amsmath,amsthm,amssymb}
\usepackage[dvipdfm]{graphicx}
\usepackage{bm}
\usepackage{tensor}
\usepackage{multicol}

\newcommand{\cK}{{\cal K}}
\newcommand{\lag}{L}

\begin{document}
\maketitle
\begin{abstract}
In this paper we propose an investing strategy based on neural
network models combined with ideas from game-theoretic probability of Shafer and Vovk.
Our proposed strategy uses parameter values of a neural network with 
the best performance until the previous round (trading day) for deciding the investment
in the current round.  We compare performance of our proposed strategy with
various strategies including a strategy based on supervised neural network models
and show that our procedure is competitive with 
other strategies.
\end{abstract}

\section{Introduction}
\label{sec:intro}

A number of researches have been conducted on
prediction of financial time series with neural networks since 
Rumelhart \cite{rumelhart} developed back propagation algorithm in 1986, 
which is the most commonly used algorithm for supervised neural network.
With this algorithm the network learns its internal structure by  updating the parameter values
when we give it training data containing inputs and outputs. We can then use the network with updated parameters 
to predict future events containing inputs the network has never encountered. 
The algorithm is applied in many fields such as robotics and image processing
and it shows a good performance in prediction of financial time series. Relevant papers on the use of
neural network to financial time series include
\cite{bernd}, \cite{hanias}, \cite{khoa} and \cite{yoon}. 

In these papers authors are concerned with the prediction of time series
and they to not pay much attention to  actual investing strategies, although the prediction
is obviously important in 
designing practical investing strategies. 
A forecast of tomorrow's price does not immediately tell us how much to invest today.
In contrast to these works, in this paper 
we directly consider investing strategies for financial time series 
based on neural network models and 
ideas from  game-theoretic probability of Shafer and Vovk (2001) \cite{glenn}.
In the game-theoretic probability established by Shafer and Vovk, 
various theorems of probability theory, such as the strong law of large numbers and the central limit theorem, 
are proved by consideration of capital processes of betting strategies in 
various games such as the coin-tossing game and the bounded forecasting game. 
In game-theoretic probability a player ``Investor'' is regarded as playing against another
player  ``Market''.
In this framework investing strategies  of Investor  play a prominent role.  
Prediction is then derived based on strong investing strategies (cf.\ defensive forecasting
in \cite{defensive}).

Recently 
in \cite{ktt-bounded}
we proposed sequential optimization of parameter values of a simple
investing strategy in multi-dimensional bounded forecasting games and showed that the resulting
strategy is easy to implement and shows a good performance in comparison to
well-known strategies such as the universal portfolio \cite{cover:1991} developed by
Thomas Cover and his collaborators. In this paper we propose 
sequential optimization of parameter values of investing strategies based on neural networks.
Neural network models give a very flexible framework for designing investing strategies.
With simulation and with some data from 
Tokyo Stock Exchange we show that
the proposed strategy shows a good performance.

The organization of this paper is as follows. In Section 
\ref{sec:sosnn} we propose sequential optimizing strategy with neural networks. 
In Section \ref{sec:compare} we present some alternative strategies for the
purpose of comparison.
In Section \ref{subsec:back-propagation} we consider an investing strategy 
using supervised neural network with back propagation algorithm. The strategy is closely related to 
and reflects existing researches on stock price prediction  with neural networks. 
In Section \ref{subsec:markovian} 
we consider Markovian proportional betting strategies, which are
much simpler than the strategies based on neural networks.
In Section \ref{sec:sim} we evaluate performances of
these strategies by Monte Carlo simulation.
In Section \ref{sec:real}
we apply these strategies to stock price data from Tokyo Stock Exchange. 
Finally  we give some concluding remarks in Section 
\ref{sec:remarks}. 

\section{Sequential optimizing strategy with neural networks}
\label{sec:sosnn}
Here we introduce the bounded forecasting game of
Shafer and Vovk \cite{glenn} in Section \ref{subsec:bounded-forecasting-game}
and network models we use in Section \ref{subsec:design-of-network}.
In Section \ref{subsec:gradient}
we specify the investing ratio 
by an unsupervised neural network and we 
propose sequential optimization of parameter values of the network. 

\subsection{Bounded forecasting game}
\label{subsec:bounded-forecasting-game}

We present the bounded forecasting game formulated by Shafer and Vovk in 2001 \cite{glenn}.
In the bounded forecasting game, Investor's capital 
at the end of round $n$ is written as $\cK_{n}$ ($n=1,2,\ldots $) 
and initial capital $\cK_{0}$ is set to be $1$. In each round Investor first announces 
the amount of money $M_n$ 
he bets ($|M_{n}| <  \cK_{n-1}$) and then Market announces her move $x_{n}\in [-1,1]$. 
$x_n$ represents the change of the price of a unit financial asset in round $n$.
The bounded forecasting 
game can be considered as an extension of the classical coin-tossing game since the bounded forecasting 
game results in the classical coin-tossing game if $x_{n}\in \{-1,1\}$. With $\cK_{n}, M_{n}$ and $x_{n}$, 
Investor's capital after round $n$ is written as $\cK_{n}=\cK_{n-1}+M_{n}x_{n}$.

The protocol of the bounded forecasting game is written as follows.

\medskip
\noindent
{\bf Protocol:}\\
\hspace*{2em}${\cK}_{0}$ $=$1.\\
\hspace*{2em}FOR $n=1,2,\dots$ :\\
\hspace*{4em}Investor announces $M_{n}\in \mathbb{R}$.\\
\hspace*{4em}Market announces $x_{n}\in [-1,1]$.\\
\hspace*{4em}${\cK}_{n}=\cK_{n-1}+M_{n}x_{n}$\\
\hspace*{2em}END FOR\\

We can rewrite Investor's capital as $\cK_{n}=\cK_{n-1}\times(1+\alpha_{n}x_{n})$, 
where $\alpha_{n}=M_n/\cK_{n-1}$ is the ratio of Investor's investment $M_n$ to his capital $\cK_{n-1}$
after round $n-1$.  We call $\alpha_n$ the investing ratio at round $n$.
We restrict 
$\alpha_{n}$ as $-1 <  \alpha_{n} <  1$ in order to prevent Investor becoming bankrupt.
Furthermore we can write $\cK_{n}$ as 
\[
\cK_{n} = \cK_{n-1}(1+\alpha_{n}x_{n})
=
\cdots
= \Pi_{k=1}^{n}(1+\alpha_{k}x_{k}).
\]
Taking the logarithm of $\cK_{n}$ we have
\begin{equation}
\label{eq:log-capital}
\log \cK_{n} = \displaystyle\sum _{k=1}^{n}\log (1+\alpha_{k} x_{k}).
\end{equation}

The behavior of Investor's capital in (\ref{eq:log-capital}) depends on the choice of 
$\alpha _{k}$.  Specifying a functional form of $\alpha_k$ is regarded  as an investing strategy. 
For example, setting $\alpha _{k}\equiv \epsilon$ to be a constant $\epsilon$ 
for all $k$ is called the $\epsilon$-strategy 
which is presented in \cite{glenn}. In  this paper we consider 
various ways to determine $\alpha_{k}$ in terms of past values $x_{k-1}, x_{k-2},\dots,$
of $x$ and and seek better $\alpha_{k}$ 
in trying to maximize the future capital $\cK_{n}$, $n>k$.

Let $\bm{u}_{k-1}=(x_{k-1}, \dots, x_{k-\lag})$ denote past $\lag$ values of $x$ and
let $\alpha_k$ depend on $\bm{u}_{k-1}$ and a parameter $\omega$: 
$\alpha_k = f(\bm{u}_{k-1}, \omega)$. Then
\[
\omega_{k-1}^* = \mathrm{argmax}\; \sum_{t=1}^{k-1} \log(1+f(\bm{u}_{t-1},\omega)x_{t})
\]
is the best parameter value until the previous round.  In our sequential optimizing
investing strategy, we use 
$\omega_{n-1}^*$ to determine the investment $M_n$ at round $n$:
\[
M_n= \cK_{n-1}\times f(\bm{u}_{n-1},\omega_{n-1}^*).
\]
For the function $f$ we employ neural network models for their flexibility, which
we describe in the next section.

\subsection{Design of the network}
\label{subsec:design-of-network}

\begin{figure}[thbp]
\vspace*{5mm}
\begin{center}
\includegraphics[width=75mm]{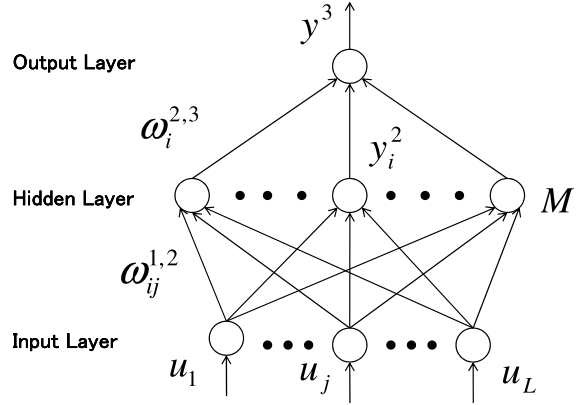}
\end{center}
\caption{Three-layered network}
\label{fig:1}
\vspace*{5mm}
\end{figure}

We construct a three-layered neural network shown in Figure
\ref{fig:1}. The input layer has $\lag $ neurons and they just
distribute the input $u_{j}\ (j=1,\dots,\lag )$ to every neuron in the
hidden layer.  Also the hidden layer has $M$ neurons and we write the
input to each neurons as $I_{i}^{2}$ which is a weighted sum of
$u_{j}$'s.

As seen  from Figure \ref{fig:1}, $I_{i}^{2}$ is obtained as
\begin{equation*}
I_{i}^{2} = \displaystyle\sum_{j=1}^{\lag }\omega_{ij}^{1,2}u_{j}, 
\end{equation*}
where $\omega_{ij}^{1,2}$ is called the weight representing the synaptic connectivity between 
the $j$th neuron in the input layer and the $i$th neuron in the hidden layer. 
Then the output of the $i$th neuron in the hidden layer is described as
\begin{equation*}  
y_{i}^{2}=\tanh (I_{i}^{2}).
\end{equation*}
As for activation function we employ hyperbolic tangent function. In a similar way, the input to 
the neuron in the output layer, which we write $I^{3}$, is obtained as
\begin{equation*}
I^{3} = \displaystyle\sum_{i=1}^{M}\omega_{i}^{2,3}y_{i}^{2},
\end{equation*}
where $\omega_{i}^{1}$ is the weight between the $i$th neuron in the hidden layer and the neuron in the output layer. 
Finally, we have
\begin{equation*}
y^{3} = \tanh (I^{3}),
\end{equation*}
which is the output of the network. In the following argument we use $y^{3}$ as an investment strategy.
Thus we can write 
\[
\alpha_{k} = y^{3}=f(\bm{u}_{k-1}, \bm{\omega}),
\]
where
\[
\bm{\omega}=(\bm{\omega}^{1,2}, \bm{\omega}^{2,3})=
\big( (\omega_{ij}^{1,2})_{i=1,\dots,M,\,j=1,\dots,\lag}, (\omega^{2,3}_i)_{i=1,\dots,M}\big).
\]
Investor's capital is written as 
\[
\cK_{n}=\cK_{n-1}(1+f(\bm{u}_{n-1}, \bm{\omega})x_{n}).
\]

We need to specify the number of inputs $\lag$ and the number of neurons $M$ in the hidden layer.
It is difficult to specify them in advance. 
We compare various choices of $\lag$ and $M$ in 
Section \ref{sec:sim} and  
Section \ref{sec:real}. 
Also in $\bm{u}_{n-1}$ we can include any input which is available
before the start of round $n$, such as moving averages of past prices, seasonal indicators or
past values of other economic time series data. 
We give further discussion on the choice of $\bm{u}_{n-1}$
in Section \ref{sec:remarks}. 

\subsection{Sequential optimizing strategy with neural networks}
\label{subsec:gradient}
In this section we propose a strategy which we call
Sequential Optimizing Strategy with Neural Networks (SOSNN).
 
We first calculate  $\bm{\omega}^*=\bm{\omega}_{n-1}^*$ that maximizes 
\begin{equation}
\label{eq:maximize}
\phi=\displaystyle\sum_{k=1}^{n-1}\log (1+f(\bm{u}_{k-1}, \bm{\omega}) x_k) .
\end{equation}
This is the best parameter values until the previous round. If Investor uses
$\alpha_n = f(\bm{u}_{n-1}, \bm{\omega}_{n-1}^*)$ as the investing ratio,
Investor's capital after round $n$ is written as 
\[
\cK_{n}=\cK_{n-1}(1+f(\bm{u}_{n-1}, \bm{\omega}_{n-1}^*)x_n).
\]

For maximization of (\ref{eq:maximize}), we employ the gradient descent method. With this method, the weight 
updating algorithm of $\omega_{i}^{2,3}$ with the parameter $\beta$ (called the learning constant) is written as 
\[
\omega _{i}^{2,3}=\omega_{i}^{2,3} + \Delta \omega_{i}^{2,3}=\omega _{i}^{2,3}+\beta \frac{\partial \phi}{\partial \omega ^{2,3}_{i}}, 
\]
where
\begin{eqnarray}
\frac{\partial \phi }{\partial \omega_{i}^{2,3}} &=& \frac{\partial \phi }{\partial f}\frac{\partial f}{\partial \omega_{i}^{2,3}}
=\displaystyle\sum_{k=1}^{n-1}\frac{\partial \phi }{\partial f}\frac{\partial f}{\partial \,{}^{k}\!I^{3}}\frac{\partial \,{}^{k}\!I^{3}}{\partial \omega _{i}^{2,3}}\nonumber \\
&=& \displaystyle\sum_{k=1}^{n-1} \frac{x_{k}}{1+ f(\bm{u}_{k-1}, \bm{\omega})x_{k}}
(1-\tanh ^{2}({}^{k}\!I^{3}))\,{}^{k}\!y_{i}^{2}\nonumber \\
&=& \displaystyle\sum_{k=1}^{n-1} \,{}^{k}\!\delta_{1} \,{}^k\!y_{i}^{2},\nonumber
\end{eqnarray}
and the left superscript $k$ to $I^3, y_i^2$ indexes the round.
Thus we obtain
\[
\Delta \omega_{i}^{2,3}=\beta \frac{\partial \phi}{\partial \omega_{i}^{2,3}}=\beta\displaystyle\sum_{k=1}^{n-1} \,{}^{k}\!{\delta_{1}}\,{}^{k}\!y_{i}^{2}.
\]
Similarly, the weight updating algorithm of $\omega_{ij}^{1,2}$ is expressed as 
\[
\omega_{ij}^{1,2}=\omega_{ij}^{1,2}+\Delta \omega_{ij}^{1,2}=\omega_{ij}^{1,2}+\beta \frac{\partial \phi}{\partial \omega_{ij}^{1,2}},
\]
where
\begin{eqnarray}
\frac{\partial \phi }{\partial \omega_{ij}^{1,2}} &=& 
\displaystyle\sum_{k=1}^{n-1}\frac{\partial \phi }{\partial \,{}^{k}\!I_{i}^{2}}
\frac{\partial \,{}^{k}\!I_{i}^{2}}{\partial \omega_{ij}^{1,2}}
=\displaystyle\sum_{k=1}^{n-1}\frac{\partial \phi }{\partial f}\frac{\partial f}{\partial\,{}^{k}\!I^{3}}\frac{\partial\,{}^{k}\!I^{3}}{\partial\,{}^{k}\!y_{i}^{2}}\frac{\partial\,{}^{k}\!y_{i}^{2}}{\partial\,{}^{k}\!I_{i}^{2}}\frac{\partial\,{}^{k}\!I_{i}^{2}}{\partial \omega_{ij}^{1,2}}\nonumber \\
&=& \displaystyle\sum_{k=1}^{n-1} \,{}^{k}\!\delta_{1}\omega_{i}^{2,3}(1-\tanh ^{2}(^{k}\!I_{i}^{2})) \, (\bm{u}_{k-1})_{j}\nonumber \\
&=& \displaystyle\sum_{k=1}^{n-1} \,{}^{k}\!\delta_{2} \, (\bm{u}_{k-1})_{j}.\nonumber
\end{eqnarray}
Thus we obtain
\[
\Delta \omega_{ij}^{1,2}=\beta \frac{\partial \phi}{\partial \omega_{ij}^{1,2}}=\beta\displaystyle\sum_{k=1}^{n-1}\,{}^{k}\!\delta_{2} (\bm{u}_{k-1})_{j}.
\]

Here we summarize the algorithm of SOSNN at round $n$.
\begin{enumerate}
\item 
Given the input vector $\bm{u}_{k-1}=(x_{k-1}, \dots, x_{k-\lag})$ $(k=1, \dots, n-1)$ and the value of $\bm{\omega}_{n-1}$, 
we first evaluate $^{k}\!I_{i}^{2} = \displaystyle\sum_{j=1}^{\lag }\omega_{ij}^{1,2} (\bm{u}_{k-1})_{j}$ and then $^{k}\!y_{i}^{2} = \tanh (^{k}\!I_{i}^{2})$. 
Also we set the learning constant $\beta$.
\item  We calculate $^{k}\!I^{3} = \displaystyle\sum_{i=1}^{M}\omega_{i}^{2,3}\,{}^{k}\!y_{i}^{2}$ and then $^{k}\!y^{3} = \tanh(^{k}\!I^{3})$ 
with $^{k}\!I_{i}^{2}$ and $^{k}\!y_{i}^{2}$ of the previous step. Then we update weight 
$\bm{\omega}$ 
with the weight updating formula $\omega_{i}^{2,3}=\omega_{i}^{2,3}+\beta\displaystyle\sum_{k=1}^{n-1}\,{}^{k}\!\delta_{1}\,{}^{k}\!y_{i}^{2}$ 
and $\omega_{ij}^{1,2} = \omega_{ij}^{1,2}+\beta\displaystyle\sum_{k=1}^{n-1}\,{}^{k}\!\delta_{2} (\bm{u}_{k-1})_{j}$.
\item  Go back to step 1 replacing the weight 
$\bm{\omega}_{n-1}$ with updated values.
\end{enumerate}
After sufficient times of iteration, $\phi$ in (\ref{eq:maximize}) converges to a local maximum 
with respect to $\omega_{ij}^{1,2}$ and $\omega_{i}^{2,3}$ and we set $\omega_{ij}^{1,2}={\omega_{ij}^{1,2}}^{*}$ and $\omega_{i}^{2,3}={\omega_{i}^{2,3}}^{*}$, 
which are elements of $\bm{\omega}_{n-1}^{*}$. Then we evaluate Investor's capital after round $n$ as 
$\cK_{n}=\cK_{n-1}(1+f(\bm{u}_{n-1}, \bm{\omega}_{n-1}^*)x_n)$. 

\section{Alternative investment strategies}
\label{sec:compare}
Here we present some strategies that are designed to be compared with SOSNN. 
In Section \ref{subsec:back-propagation} we present a strategy with back-propagating neural network. 
The advantage of back-propagating neural network is its predictive ability due to ``learning'' as previous researches show. 
In Section \ref{subsec:markovian} we show some sequential optimizing strategies that use rather simple function for $f$ 
than SOSNN does. 

\subsection{Optimizing strategy with back propagation}
\label{subsec:back-propagation}
In this section we consider a supervised neural network and its optimization by back propagation.
We call the strategy NNBP.  
It decides the betting ratio by predicting actual up-and-downs of stock prices and can be regarded as 
incorporating existing researches on stock price prediction. Thus it is suitable as an alternative to SOSNN. 

For supervised network, we train the network
with the data from a training period, 
obtain the best value of the parameters for the training period and then use it for 
the investing period. These two periods are distinct.
For the training period we need to specify the desired output (target) $T_{k}$ of the
network for each day $k$.  We propose to specify the target by the direction of 
Market's current price movement $x_{k}$.
Thus we set
\[
T_{k}=
\begin{cases}
+1 & x_{k}> 0\\
0 & x_{k}=0\\
-1 & x_{k}<0
\end{cases}
.\]
Note that this $T_{k}$ is the best investing ratio 
if Investor could use the current movement $x_{k}$ of Market
for his investment. 
Therefore it is natural to use $T_{k}$ as the target value
for investing strategies. We keep on updating $\bm{\omega}_{k}$ by
cycling through the input-output pairs of the days of the training period 
and finally obtain $\bm{\omega}^{*}$  after sufficient times of iteration. 

Throughout the investing period we use $\bm{\omega}^{*}$ and 
Investor's capital after round $n$ in the investing period is expressed as 
\[
\cK_{n}=\cK_{n-1}(1+f(\bm{u}_{n-1}, \bm{\omega}^{*})x_{n}).
\]

Back propagation is an algorithm which updates weights ${}^{k}\!\omega_{ij}^{1,2}$ and ${}^{k}\!\omega_{i}^{2,3}$ so that the error function 
\[
E_{k}=\frac{1}{2}(T_{k}- \,{}^{k}\!y^{3})^{2}
\] 
decreases, where $T_{k}$ is the desired output of the network and $^{k}\!y^{3}$ is the actual output of the network. 
The weight ${}^{k}\!\omega_{i}^{2,3}$ of day $k$ is renewed to the weight
${}^{k+1}\!\omega_{i}^{2,3}$ of day $k+1$ as
\begin{eqnarray}
{}^{k+1}\!\omega_{i}^{2,3} &=& {}^{k}\!\omega_{i}^{2,3} + \Delta{}^{k}\!\omega_{i}^{2,3}={}^{k}\!\omega_{i}^{2,3}-\beta\frac{\partial E_{k}}{\partial{}^{k}\!\omega_{i}^{2,3}}
={}^{k}\!\omega_{i}^{2,3}-\beta\frac{\partial E_{k}}{\partial\,{}^{k}\!I^{3}}\frac{\partial\,{}^{k}\!I^{3}}{\partial{}^{k}\!\omega_{i}^{2,3}}\nonumber \\
&=& {}^{k}\!\omega_{i}^{2,3}-\beta\,{}^{k}\!\epsilon^{1}\,{}^{k}\!y_{i}^{2},\nonumber
\end{eqnarray}
where
\[
^{k}\!\epsilon^{1} = \frac{\partial E_{k}}{\partial\,{}^{k}\!I^{3}}=\frac{\partial E_{k}}{\partial\,{}^{k}\!y^{3}}\frac{\partial\,{}^{k}\!y^{3}}{\partial\,{}^{k}\!I^{3}}
= -(T_{k}-^{k}\!\!y^{3})(1-\tanh^{2}(^{k}\!I^{3})).
\]
Also weight ${}^{k}\!\omega_{ij}^{1,2}$ is renewed as
\begin{eqnarray}
{}^{k+1}\!\omega_{ij}^{1,2} &=& {}^{k}\!\omega_{ij}^{1,2}+\Delta{}^{k}\!\omega_{ij}^{1,2}={}^{k}\!\omega_{ij}^{1,2}-\beta\frac{\partial E_{k}}{{}^{k}\!\omega_{ij}^{1,2}}
={}^{k}\!\omega_{ij}^{1,2}-\beta\frac{\partial E_{k}}{\partial\,{}^{k}\!I^{3}}\frac{\partial\,{}^{k}\!I^{3}}{\partial\,{}^{k}\!y_{i}^{2}}\frac{\partial\,{}^{k}\!y_{i}^{2}}{\partial\,{}^{k}\!I_{i}^{2}}\frac{\partial\,{}^{k}\!I_{i}^{2}}{\partial{}^{k}\!\omega_{ij}^{1,2}}\nonumber \\
&=& {}^{k}\!\omega_{ij}^{1,2}-\beta\,{}^{k}\!\epsilon_{i}^{2} \,(\tilde{\bm{u}}_{k-1})_{j},\nonumber 
\end{eqnarray}
where
\[
^{k}\!\epsilon_{i}^{2} = \frac{\partial E_{k}}{\partial\,{}^{k}\!I^{3}}\frac{\partial\,{}^{k}\!I^{3}}{\partial\,{}^{k}\!y_{i}^{2}}\frac{\partial\,{}^{k}\!y_{i}^{2}}{\partial\,{}^{k}\!I_{i}^{2}}
=\,{}^{k}\!\epsilon^{1}\omega_{i}^{2,3}(1-\tanh^{2}(^{k}\!I_{i}^{2})) . 
\]
At the end of each step we calculate the training error defined as 
\begin{equation}
\label{eq:te}
\mbox{training error}=\frac{1}{2m}\sum_{k=1}^m (T_k-{}^{k}\!y^{3})^2=\frac{1}{m}\sum_{k=1}^m E_k,
\end{equation}
where $m$ is the length 
of the training period.  We end the iteration 
when the the training error becomes smaller than the threshold $\mu$, which is set sufficiently small. 

Here let us summarize the algorithm of NNBP in the training period.
\begin{enumerate}
\item We set $k=1$.
\item Given the input vector $\bm{u}_{k-1}=(x_{k-1}, \dots, x_{k-\lag})$ and the value of 
$\bm{\omega}_{k}$, 
we first evaluate $^{k}\!I_{i}^{2}
=\displaystyle\sum_{j=1}^{\lag }{}^{k}\!\omega_{ij}^{1,2}(\bm{u}_{k-1})_{j}$ 
and then $^{k}\!y_{i}^{2}=\tanh(^{k}\!I_{i}^{2})$. Also we set the learning constant $\beta$.
\item We calculate $^{k}\!I^{3} = \displaystyle\sum_{i=1}^{M}{}^{k}\!\omega_{i}^{2,3} 
\,{}^{k}\!y_{i}^{2}$ and then 
$^{k}\!y^{3}=\tanh(^{k}\!I^{3})$ with $^{k}\!I_{i}^{2}$ and $^{k}\!y_{i}^{2}$ of the previous step. 
Then we update weight 
$\bm{\omega}_{k}$
with the weight updating formula 
${}^{k+1}\!\omega_{i}^{2,3}={}^{k}\!\omega_{i}^{2,3}-\beta\,{}^{k}\!\epsilon^{1}\,{}^{k}\!y_{i}^{2}$ 
and ${}^{k+1}\!\omega_{ij}^{1,2}={}^{k}\!\omega_{ij}^{1,2}-\beta\,{}^{k}\!\epsilon_{i}^{2}(\bm{u}_{k-1})_{j}$.
\item Go back to step 2 setting $k+1\leftarrow k$ and $\bm{\omega}_{k+1}\leftarrow \bm{\omega}_{k}$ 
while $1\le k\le m$. When $k=m$ we set 
$k=1$ and $\bm{\omega}_{1}\leftarrow \bm{\omega}_{m+1}$ and continue the algorithm 
until the training error becomes less than $\mu$. 
\end{enumerate}

\subsection{Markovian proportional betting strategies}
\label{subsec:markovian}
In this section we present some sequential optimizing strategies that are rather simple compared to strategies with 
neural network in Section \ref{sec:sosnn} and Section \ref{subsec:back-propagation}.
The strategies of this section are generalizations of Markovian strategy in \cite{tkt-multistep}
for coin-tossing games to bounded forecasting games. We present these simple strategies for 
comparison with SOSNN and observe how complexity in function $f$ 
increases or decreases Investor's capital processes in numerical examples in later sections.

Consider maximizing the logarithm of Investor's capital in (\ref{eq:log-capital}):
\[
\log \cK_{n} = \displaystyle\sum _{k=1}^{n}\log (1+\alpha_{k} x_{k}).
\]
We first consider the following simple strategy of \cite{ktt-bounded}
in which we use $\alpha_{n}=\alpha_{n-1}^{*}$, where 
\[
\alpha_{n-1}^{*}=\mathrm{argmax}\; \displaystyle\Pi_{k=1}^{n-1}(1+\alpha x_{k}).
\]
In this paper we denote this strategy by MKV0.

As a generalization of MKV0 consider using different investing ratios
depending on whether the price went up or down on the previous day.  Let 
$\alpha_{k}=\alpha_{k}^{+}$ when $x_{k-1}$ 
was positive and $\alpha_{k}=\alpha_{k}^{-}$ when it was negative. 
We denote this strategy by MKV1.
In the betting on the $n$th day 
we use $\alpha_{n}^{+}={\alpha_{n-1}^{+}}^{*}$ and $\alpha_{n}^{-}={\alpha_{n-1}^{-}}^{*}$, where 
\[
({\alpha_{n-1}^{+}}^{*}, {\alpha_{n-1}^{-}}^{*}) =\mathrm{argmax}\; \displaystyle\Pi_{k=1}^{n-1}(1+f(\bm{u}_{k-1}, \alpha^{+}, \alpha^{-}) x_{k}),
\]
$\bm{u}_{k-1}=(x_{k-1})$ and 
\[
f(\bm{u}_{k-1}, \alpha^{+}, \alpha^{-})=\alpha^{+}\mathrm{I}_{\{ x_{k-1}\ge 0\}}+\alpha^{-}\mathrm{I}_{\{ x_{k-1}< 0\}}.
\]
Here $\mathrm{I}_{\{\cdot\}}$ denotes the indicator function of the event in $\{\cdot\}$.
The capital process of MKV1 is written in the form of (\ref{eq:log-capital}) as
\[
\log \cK_{n} = \displaystyle\sum _{k=1}^{n}\log (1+f(\bm{u}_{k-1}, {\alpha_{k-1}^{+}}^{*}, {\alpha_{k-1}^{-}}^{*})x_{k}).
\]

We can further generalize this strategy considering price movements of past two days.
Let $\bm{u}_{k-1}=(x_{k-1}, x_{k-2})$  and let
\begin{align*}
f(\bm{u}_{k-1}, \alpha^{++}, \alpha^{+-}, \alpha^{-+}, \alpha^{--})
&= \alpha^{++} \mathrm{I}_{\{x_{k-2}\ge 0,\; x_{k-1}\ge 0\}} 
+\alpha^{+-} \mathrm{I}_{\{x_{k-2}\ge 0, \; x_{k-1}< 0\}}  \\
& \qquad +\alpha^{-+} \mathrm{I}_{\{x_{k-2}< 0, \; x_{k-1}\ge 0\}} 
+\alpha^{--} \mathrm{I}_{\{x_{k-2}< , x_{k-1}< 0\}}.
\end{align*}
We denote this strategy by MKV2.

We will compare performances of the above Markovian proportional betting strategies
with strategies based on neural networks in the following sections.

\section{Simulation with linear models}
\label{sec:sim}

In this section we give some simulation results for strategies shown in Section
\ref{sec:sosnn} and Section \ref{sec:compare}. 
We use two linear time series models to confirm the behavior of presented strategies.
Linear time series data are generated from the Box-Jenkins family \cite{box}, 
autoregressive model of order 1 (AR(1)) and autoregressive moving average model of order 2 and 1 (ARMA(2,1))  
having the same parameter values as in \cite{zhang}. AR(1) data are generated as 
\begin{equation}
\label{eq:ar1data}
x_{n}=0.6 x_{n-1}+\epsilon_{n}
\end{equation}
and ARMA(2,1) data are generated as 
\begin{equation}
\label{eq:armadata}
x_{n}=0.6 x_{n-1}+0.3 x_{n-2}+\epsilon_{n}-0.5\epsilon_{n-1},
\end{equation}
where we set $\epsilon_{n}\sim N(0,1)$. 
After the series is generated, we divide each value by the maximum absolute value to normalize the data to the admissible range $[-1,1]$. 

Here we discuss some details on calculation of each strategy. 
First we set the initial values of elements of $\bm{\omega}$ as random numbers in $[-0.1, 0.1]$. 
In SOSNN, we use the first $20$ values of $x_{n}$ as 
initial values and the iteration process in gradient descent method is proceeded 
until $|\Delta\omega_{ij}^{1,2}|<10^{-4}$ and $|\Delta\omega_{i}^{2,3}|<10^{-4}$ with the upper bound of $10^{4}$ steps. 
As for the learning constant $\beta$, we use learning-rate annealing schedules which appear in Section 3.13 of 
\cite{simon}. 
With annealing schedule called the search-then-converge schedule \cite{darken}
we put $\beta$ at the $n$th step of iteration as  
\[
\beta (n)=\frac{\beta_{0}}{1+( n/\tau )},
\]
where $\beta_{0}$ and $\tau$ are constants and we set $\beta_{0}=1.0$ and $\tau=5.0$. 
In NNBP, we train the network with five different training sets of $300$ observations generated by (\ref{eq:ar1data}) and (\ref{eq:armadata}). 
We continue cycling through the training set until the training error becomes less than $\mu$ and we set $\mu=10^{-2}$ 
with the upper bound of $6\times 10^{5}$ steps. 
Also we check the fit of the network to the data by means of the training error for some different values of $\beta$, $\lag$ and $M$. 
In Markovian strategies, we again use the first $20$ values of $x_{n}$ as initial values. 
We also adjust the data so that the betting is conducted  on the same data regardless of $\lag$ in SOSNN and NNBP 
or different number of inputs among Markovian strategies. 

In Table \ref{sim} we summarize the results of SOSNN, NNBP, MKV0, MKV1 and MKV2 under AR(1) and ARMA(2,1). 
The values presented are averages of results for five different simulation runs of (\ref{eq:ar1data}) and (\ref{eq:armadata}).
As for SOSNN, we simulate fifty cases (combinations of $\lag=1, \dots, 5$ and $M=1, \dots, 10$),  
but only report the cases of $\lag=1, 2, 3$ and some choices  of $M$ because the purpose of the simulation 
is to test whether $\lag=1$ works better than other choices of $\lag$ under AR(1) and $\lag=2$ works better under ARMA(2,1). 
For NNBP we only report the result for one case since fitting the parameters to the data is quite a  difficult task due to the 
characteristic of desired output (target). Also once we obtain the value of $\beta$, $\lag$ and $M$ with training error less than the threshold $\mu=10^{-2}$, 
we find that the network has successfully learned the input-output relationship and we do not test  other choices of the above parameters.  (See Appendix for more detail.) 
We set $\beta=0.07$, $\lag=12$ and $M=30$ in simulation with AR(1) model and $\beta=0.08$, $\lag=15$ and $M=40$ in simulation with ARMA(2,1) model. 

Investor's capital process for each choice of $\lag$ and $M$ in SOSNN, NNBP and 
each Markovian strategy is shown in three rows, corresponding to rounds $100$, $200$, $300$ of  the betting 
(without the initial 20 rounds in SOSNN and Markovian strategies). 
The fourth row of each result for NNBP shows the training error after learning in the training period. 
The best value among the choices  of $\lag$ and $M$ in SOSNN 
or among each Markovian strategy is written in bold and marked with an asterisk and
the second best value is also written in bold and marked with two asterisks. 
Also calculation results written with ``---'' are cases in that simulation did not 
give proper values for some reasons. 

\begin{table}[htbp]
\tiny
\caption{Log capital processes of SOSNN, NNBP, MKV0, MKV1, MKV2 under AR(1) and ARMA(2,1)}
\label{sim}
\setlength{\tabcolsep}{2pt}
\begin{center}
\begin{tabular}{ccccccccc}
\hline
\hline
&&&&&&&&\\
\multicolumn{3}{c}{\underline{\bf AR(1) model}}&&&&&&\\
&\multicolumn{8}{c}{\underline{\bf SOSNN}}\\
&&&&&&&&\\
$\lag\verb|\|M$&$1$&$2$&$3$&$4$&$5$&$6$&$7$&$8$\\
&&&&&&&&\\
&$9.890$&$9.168$&$\bm{10.021}^*$&$10.007$&$9.982$&$9.559$&$\bm{10.009}^{**}$&$9.821$\\
$1$&$18.285$&$17.436$&$17.241$&$17.709$&$\bm{18.480}^*$&$17.542$&$\bm{18.260}^{**}$&$16.991$\\
&$30.151$&$23.574$&$29.465$&$29.949$&$\bm{32.483}^*$&$28.941$&---&---\\
&&&&&&&&\\
&$7.476$&$7.132$&$7.864$&$8.004$&$7.267$&$8.738$&$6.009$&$7.771$\\
$2$&$14.983$&$12.710$&$15.518$&$17.350$&$13.914$&$13.614$&---&$12.016$\\
&$26.211$&$25.785$&$26.090$&$\bm{32.144}^{**}$&$21.981$&$23.137$&---&$19.492$\\
&&&&&&&&\\
&$8.128$&$9.084$&$5.427$&---&$2.922$&$7.291$&$5.209$&$5.989$\\
$3$&$13.934$&$12.242$&$11.166$&---&$8.607$&$11.539$&$8.311$&$10.994$\\
&$23.232$&$19.889$&$20.013$&---&$16.330$&$18.807$&$16.055$&$20.388$\\
&&&&&&&&\\
&\multicolumn{2}{c}{\underline{\bf NNBP}}&\multicolumn{2}{c}{\underline{\bf MKV0}}&\multicolumn{2}{c}{\underline{\bf MKV1}}&\multicolumn{2}{c}{\underline{\bf MKV2}}\\
&&&&&&&&\\
&\multicolumn{2}{c}{$10.118$}&\multicolumn{2}{c}{$-1.175$}&\multicolumn{2}{c}{$\bm{7.831}^*$}&\multicolumn{2}{c}{$6.517$}\\
&\multicolumn{2}{c}{$11.921$}&\multicolumn{2}{c}{$-0.800$}&\multicolumn{2}{c}{$\bm{16.974}^*$}&\multicolumn{2}{c}{$15.452$}\\
&\multicolumn{2}{c}{$24.323$}&\multicolumn{2}{c}{$-1.647$}&\multicolumn{2}{c}{$\bm{32.392}^*$}&\multicolumn{2}{c}{$30.875$}\\
&\multicolumn{2}{c}{$(5.28\times 10^{-3})$}&&&&&\\
&&&&&&&\\
\hline
&&&&&&&&\\
\multicolumn{3}{c}{\underline{\bf ARMA(2,1) model}}&&&&&&\\
&\multicolumn{8}{c}{\underline{\bf SOSNN}}\\
&&&&&&&&\\
$\lag\verb|\|M$&$1$&$2$&$3$&$4$&$5$&$6$&$7$&$8$\\
&&&&&&&&\\
&$4.985$&$5.292$&$4.350$&$4.532$&$4.052$&$3.563$&$3.408$&$1.522$\\
$1$&$10.234$&$11.108$&$9.905$&$8.793$&$10.012$&$8.411$&$8.029$&$4.746$\\
&$11.666$&$11.460$&$10.024$&$9.924$&$11.781$&$8.016$&---&$5.882$\\
&&&&&&&&\\
&$8.518$&$10.287$&$\bm{11.052}^{**}$&$\bm{11.567}^*$&$9.147$&$8.483$&$6.915$&$5.579$\\
$2$&$18.474$&$20.177$&$\bm{21.458}^*$&$\bm{20.768}^{**}$&$13.030$&$12.042$&---&$13.573$\\
&$16.818$&$\bm{25.167}^*$&$24.979$&$\bm{25.114}^{**}$&$17.538$&$15.074$&---&$16.846$\\
&&&&&&&&\\
&$8.511$&$10.490$&$7.344$&$7.883$&$7.409$&$7.445$&$7.011$&$5.772$\\
$3$&$15.401$&$18.241$&$18.120$&$15.697$&---&$12.395$&---&$11.961$\\
&$18.047$&$24.904$&$23.362$&$21.280$&---&$15.930$&---&$14.185$\\
&&&&&&&&\\
&\multicolumn{2}{c}{\underline{\bf NNBP}}&\multicolumn{2}{c}{\underline{\bf MKV0}}&\multicolumn{2}{c}{\underline{\bf MKV1}}&\multicolumn{2}{c}{\underline{\bf MKV2}}\\
&&&&&&&&\\
&\multicolumn{2}{c}{$7.813$}&\multicolumn{2}{c}{$-0.729$}&\multicolumn{2}{c}{$3.160$}&\multicolumn{2}{c}{$\bm{7.566}^*$}\\
&\multicolumn{2}{c}{$15.420$}&\multicolumn{2}{c}{$-0.132$}&\multicolumn{2}{c}{$10.483$}&\multicolumn{2}{c}{$\bm{17.619}^*$}\\
&\multicolumn{2}{c}{$25.819$}&\multicolumn{2}{c}{$-2.422$}&\multicolumn{2}{c}{$13.375$}&\multicolumn{2}{c}{$\bm{22.911}^*$}\\
&\multicolumn{2}{c}{$(2.04\times 10^{-2})$}&&&&&\\
&&&&&&&\\
\hline
\hline
\end{tabular}
\end{center}
\end{table}

Notice that SOSNN whose betting ratio $f$ is specified by a complex function gives better performance 
than rather simple Markovian strategies both under AR(1) and ARMA(2,1). 
Also the result that NNBP gives capital processes which are competitive with other strategies shows that 
the network has successfully learned the input-output relationship in the training period. 

\section{Comparison of performances with some stock price data}
\label{sec:real}
In this section we present numerical examples calculated with the stock price data 
of three Japanese companies SONY, Nomura Holdings and NTT listed on the first section of the Tokyo Stock Exchange 
for comparing betting strategies presented in Section \ref{sec:sosnn} and Section \ref{sec:compare}. 
The investing period (without days used for initial values) is $300$ days 
from March 1st in 2007 to June 19th in 2008 for all strategies and 
the training period in NNBP is $300$ days from December 1st in 2005 to February 20th in 2007. We use a shorter training period 
than those in previous researches,  
because longer periods resulted in poor fitting. 

The data for $300$ days from December 1st in 2005 to February 20th in 2007 is used for input normalization 
of $x_{n}$ to $[-1, 1]$, which is conducted according to the method shown 
in \cite{azoff} and the procedure is as follows. For the data of daily closing prices in the above period, we first obtain 
the maximum value of absolute daily movements and then divide daily movements 
in the investing period by that maximum value. 
In case $x_{n}\le -1.0$ or $x_{n}\ge 1.0$ we put $x_{n}=-1.0$ or $x_{n}=1.0$. 
Thus we obtain $x_{n}$ in $[-1, 1]$ and we use them for inputs of the neural network. 
We tried periods of various lengths for normalization and decided to choose a relatively 
short period to avoid Investor's capital processes 
staying almost constant.
Also in NNBP we used $\beta=0.07$, $\lag=12$ and $M=90$ for SONY, $\beta=0.07$, $\lag=15$ and $M=100$ for Nomura 
and $\beta=0.03$, $\lag=15$ and $M=120$ for NTT with upper bound of $10^{5}$ iteration steps. 
Other details of  calculation are the same as in Section \ref{sec:sim}. 
We report the results in Table \ref{real}.

\begin{table}[htbp]
\tiny
\caption{Log capital process of SOSNN, NNBP, MKV0, MKV1 and MKV2 for TSE stocks}
\label{real}
\setlength{\tabcolsep}{2pt}
\begin{center}
\begin{tabular}{cccccccc}
\hline
\hline
&&&&&&&\\
\multicolumn{2}{c}{\underline{{\bf SONY}}}&&&&&&\\
&\multicolumn{7}{c}{\underline{\bf SOSNN}}\\
&&&&&&&\\
$\lag\verb|\|M$&$1$&$2$&$4$&$5$&$7$&$8$&$9$\\
&&&&&&&\\
&$-0.339$&$-0.292$&$\bm{0.144}^{**}$&$-0.832$&$-0.964$&$-0.896$&$0.082$\\
$1$&$0.572$&$0.520$&$\bm{1.438}^{**}$&$0.401$&$0.383$&$-0.433$&${\bm{2.012}}^*$\\
&$-0.153$&$-0.220$&$\bm{0.461}^{**}$&$-0.633$&$-0.762$&$-1.283$&${\bm{0.582}}^*$\\
&&&&&&&\\
&$-0.329$&$0.163$&$-1.273$&$-0.675$&$-0.349$&$-2.056$&$-2.541$\\
$2$&$0.000$&$-0.260$&$-1.825$&$-1.420$&$-0.072$&$-1.100$&$-1.256$\\
&$-0.565$&$-0.412$&$-2.274$&$-2.006$&$-0.980$&$-2.490$&$-1.981$\\
&&&&&&&\\
&$-0.230$&$-1.448$&$-1.402$&$-1.232$&$\bm{0.397}^*$&$-0.627$&$-1.404$\\
$3$&$-0.309$&$-1.823$&$-0.438$&$-0.695$&$0.970$&$-0.599$&$0.281$\\
&$-0.307$&$-2.237$&$-1.333$&$-1.843$&$-0.212$&$-1.827$&$-2.749$\\
&&&&&&&\\
\multicolumn{2}{c}{\underline{\bf NNBP}}&\multicolumn{2}{c}{\underline{\bf MKV0}}&\multicolumn{2}{c}{\underline{\bf MKV1}}&\multicolumn{2}{c}{\underline{\bf MKV2}}\\
&&&&&&&\\
\multicolumn{2}{c}{$-1.039$}&\multicolumn{2}{c}{$-0.578$}&\multicolumn{2}{c}{$\bm{-0.459}^*$}&\multicolumn{2}{c}{$-1.285$}\\
\multicolumn{2}{c}{$-2.557$}&\multicolumn{2}{c}{$-0.979$}&\multicolumn{2}{c}{$\bm{1.414}^*$}&\multicolumn{2}{c}{$-0.482$}\\
\multicolumn{2}{c}{$-3.837$}&\multicolumn{2}{c}{$-1.260$}&\multicolumn{2}{c}{$\bm{-0.212}^*$}&\multicolumn{2}{c}{$-2.297$}\\
\multicolumn{2}{c}{$(3.67\times 10^{-2})$}&&&&&&\\
&&&&&&&\\
\hline
&&&&&&&\\
\multicolumn{2}{c}{\underline{{\bf Nomura}}}&&&&&&\\
&\multicolumn{7}{c}{\underline{\bf SOSNN}}\\
&&&&&&&\\
$\lag\verb|\|M$&$1$&$2$&$4$&$5$&$7$&$8$&$9$\\
&&&&&&&\\
&$0.200$&$-0.212$&$-1.309$&$-0.479$&$\bm{0.839}^*$&$\bm{0.679}^{**}$&$-0.726$\\
$1$&$\bm{1.193}^*$&$0.754$&$-2.417$&$-0.370$&$-2.650$&$0.005$&$-1.212$\\
&$-3.326$&$-2.888$&$-2.333$&$\bm{0.581}^{**}$&$-6.938$&$-4.979$&$-0.504$\\
&&&&&&&\\
&$-0.819$&$0.338$&$-4.148$&$0.229$&$-4.787$&$-1.662$&$-0.754$\\
$2$&$-0.969$&$\bm{1.030}^{**}$&$-4.920$&$-0.007$&$-11.478$&$-3.385$&$-10.046$\\
&$-0.202$&$\bm{1.127}^*$&$-4.941$&$-7.003$&$-18.795$&$-10.897$&$-22.861$\\
&&&&&&&\\
&$-1.066$&$-2.076$&$-1.458$&$-0.389$&$-2.926$&$-1.783$&$-2.451$\\
$3$&$-1.111$&$1.002$&$-5.307$&$0.198$&$-2.897$&$-5.254$&$-9.595$\\
&$-3.672$&$-1.599$&$-10.570$&$-3.885$&$-0.621$&$-8.420$&$-14.021$\\
&&&&&&&\\
\multicolumn{2}{c}{\underline{\bf NNBP}}&\multicolumn{2}{c}{\underline{\bf MKV0}}&\multicolumn{2}{c}{\underline{\bf MKV1}}&\multicolumn{2}{c}{\underline{\bf MKV2}}\\
&&&&&&&\\
\multicolumn{2}{c}{$-0.743$}&\multicolumn{2}{c}{$\bm{-0.883}^*$}&\multicolumn{2}{c}{$-1.911$}&\multicolumn{2}{c}{$-3.789$}\\
\multicolumn{2}{c}{$-8.087$}&\multicolumn{2}{c}{$\bm{-0.753}^*$}&\multicolumn{2}{c}{$-1.354$}&\multicolumn{2}{c}{$-4.970$}\\
\multicolumn{2}{c}{$-16.051$}&\multicolumn{2}{c}{$\bm{-1.390}^*$}&\multicolumn{2}{c}{$-1.952$}&\multicolumn{2}{c}{$-6.410$}\\
\multicolumn{2}{c}{$(2.03\times 10^{-2})$}&&&&&&\\
&&&&&&&\\
\hline
&&&&&&&\\
\multicolumn{2}{c}{\underline{{\bf NTT}}}&&&&&&\\
&\multicolumn{7}{c}{\underline{\bf SOSNN}}\\
&&&&&&&\\
$\lag\verb|\|M$&$1$&$2$&$4$&$5$&$7$&$8$&$9$\\
&&&&&&&\\
&$-0.674$&$-0.365$&$-1.824$&$\bm{0.880}^{**}$&$-5.673$&$\bm{0.888}^*$&$-4.757$\\
$1$&$-0.825$&$-0.411$&$\bm{-0.246}^*$&$-4.248$&$-4.472$&$-7.931$&$-4.386$\\
&$-1.115$&---&---&---&---&---&---\\
&&&&&&&\\
&$-1.269$&$-6.049$&$-5.476$&$-2.769$&$-7.899$&$-6.606$&$-4.134$\\
$2$&$-1.175$&$-8.333$&$-9.581$&$-5.377$&$-11.071$&$-9.206$&$-5.250$\\
&$\bm{-0.498}^{**}$&$-10.660$&---&$-7.137$&---&$-13.900$&---\\
&&&&&&&\\
&$-0.377$&$-3.131$&$-2.192$&$-4.449$&$-1.670$&$-10.896$&$-8.633$\\
$3$&$\bm{-0.431}^{**}$&$-5.861$&$-1.366$&$-9.349$&$-16.990$&$-10.811$&$-12.315$\\
&$\bm{0.092}^*$&$-5.463$&---&$-14.584$&---&---&$-15.456$\\
&&&&&&&\\
\multicolumn{2}{c}{\underline{\bf NNBP}}&\multicolumn{2}{c}{\underline{\bf MKV0}}&\multicolumn{2}{c}{\underline{\bf MKV1}}&\multicolumn{2}{c}{\underline{\bf MKV2}}\\
&&&&&&&\\
\multicolumn{2}{c}{$-4.155$}&\multicolumn{2}{c}{$\bm{-0.902}^*$}&\multicolumn{2}{c}{$-2.048$}&\multicolumn{2}{c}{$-4.788$}\\
\multicolumn{2}{c}{$-3.161$}&\multicolumn{2}{c}{$\bm{-1.272}^*$}&\multicolumn{2}{c}{$-2.642$}&\multicolumn{2}{c}{$-6.807$}\\
\multicolumn{2}{c}{$-5.669$}&\multicolumn{2}{c}{$\bm{-1.566}^*$}&\multicolumn{2}{c}{$-3.391$}&\multicolumn{2}{c}{$-8.687$}\\
\multicolumn{2}{c}{$(3.73\times 10^{-2})$}&&&&&&\\
&&&&&&&\\
\hline
\hline
\end{tabular}
\end{center}
\end{table}

\begin{figure}[htpb]
\begin{minipage}{0.5\hsize}
\begin{center}
\includegraphics[width=70mm]{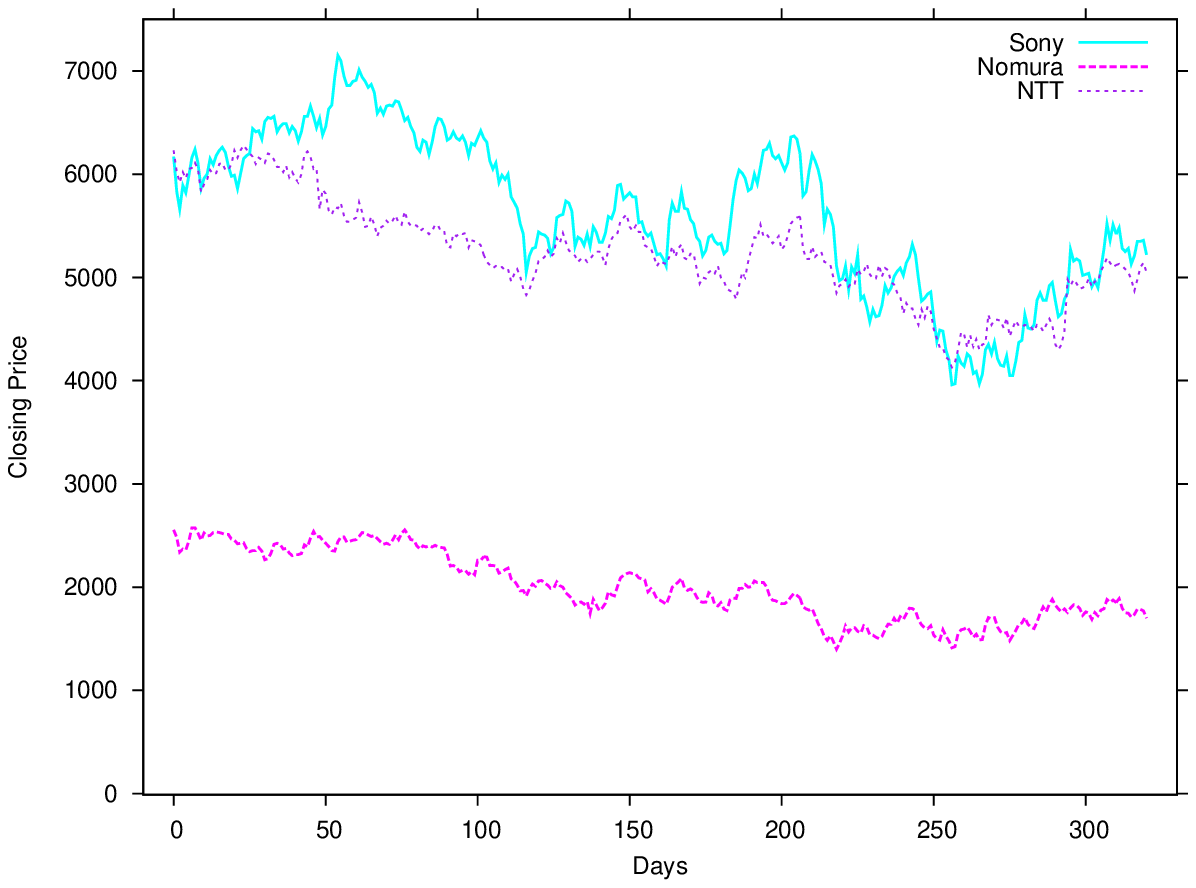}
\caption{Closing prices}
\label{fig:2}
\end{center}
\end{minipage}
\begin{minipage}{0.5\hsize}
\begin{center}
\includegraphics[width=70mm]{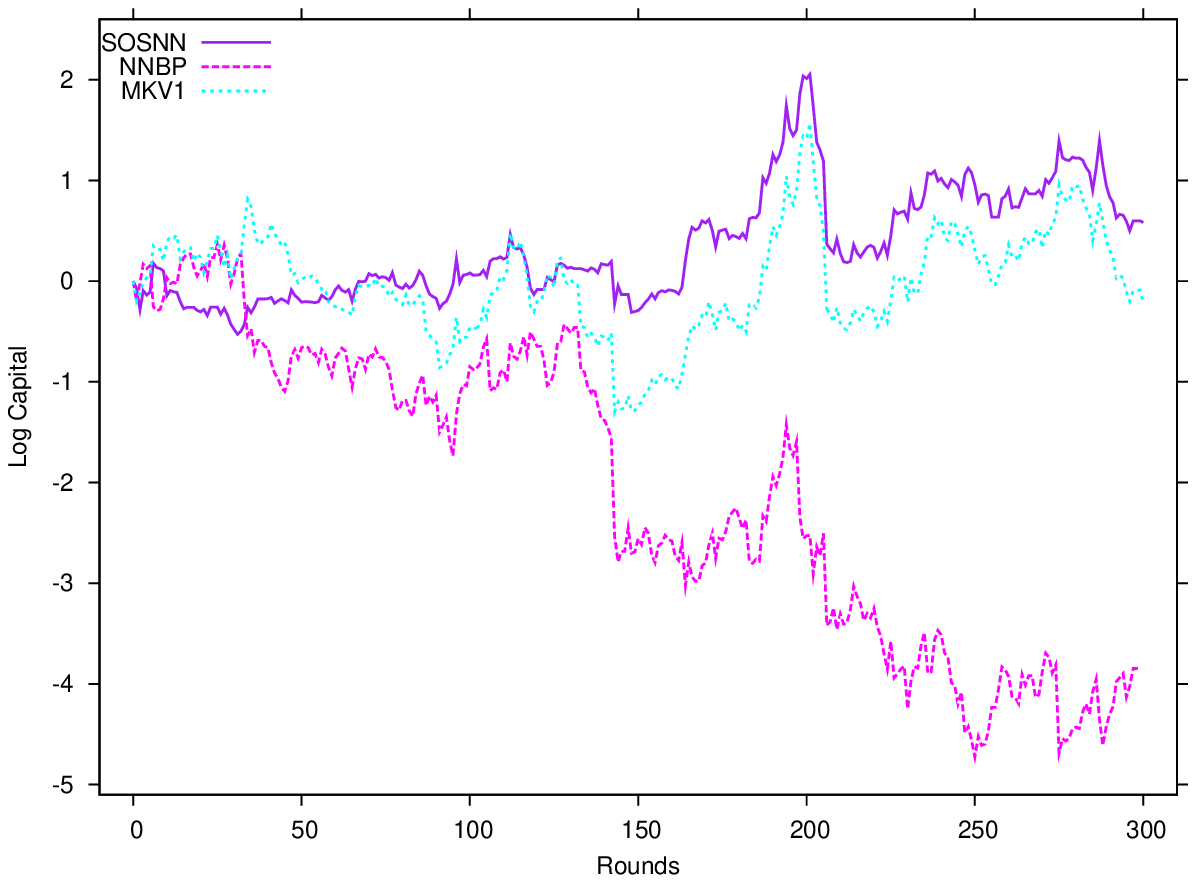}
\caption{SONY}
\label{fig:3}
\end{center}
\end{minipage}
\end{figure}

\begin{figure}[htpb]
\begin{minipage}{0.5\hsize}
\begin{center}
\includegraphics[width=70mm]{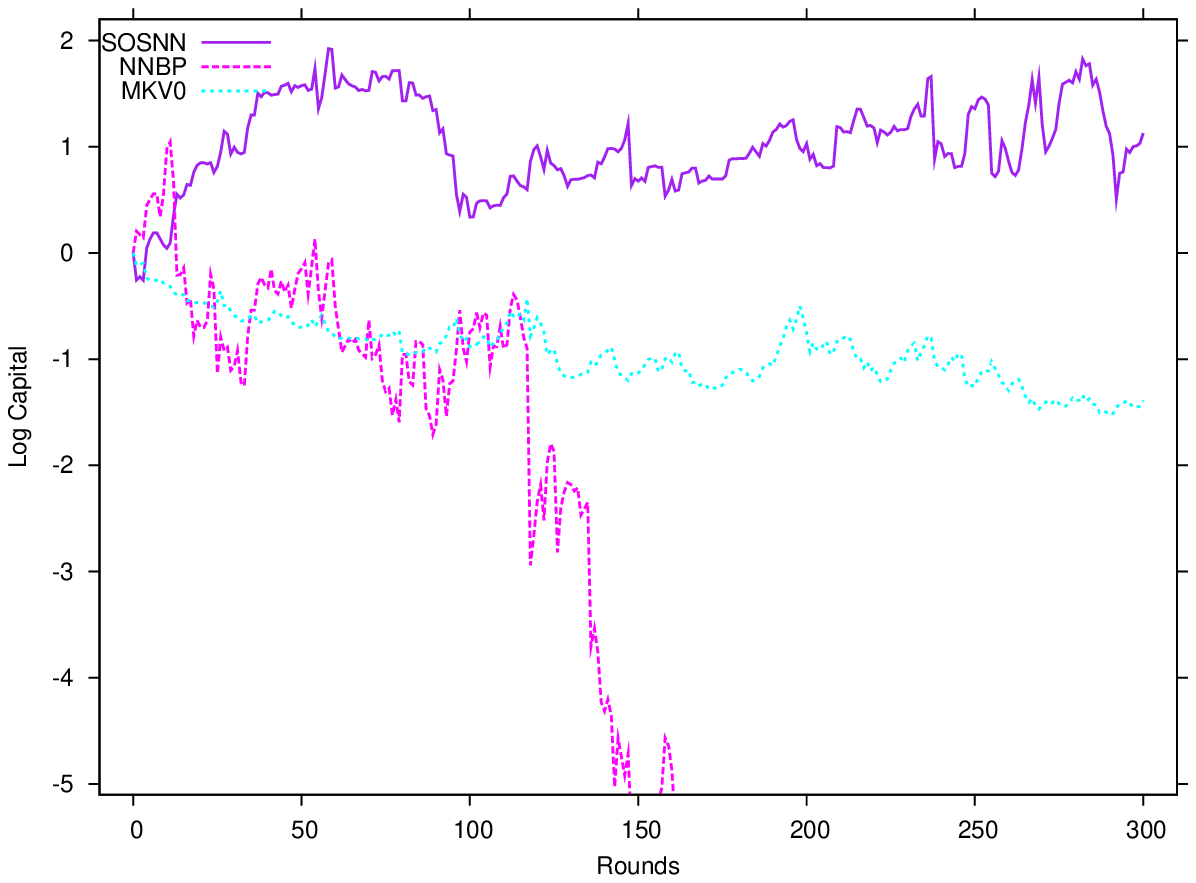}
\caption{Nomura}
\label{fig:4}
\end{center}
\end{minipage}
\begin{minipage}{0.5\hsize}
\begin{center}
\includegraphics[width=70mm]{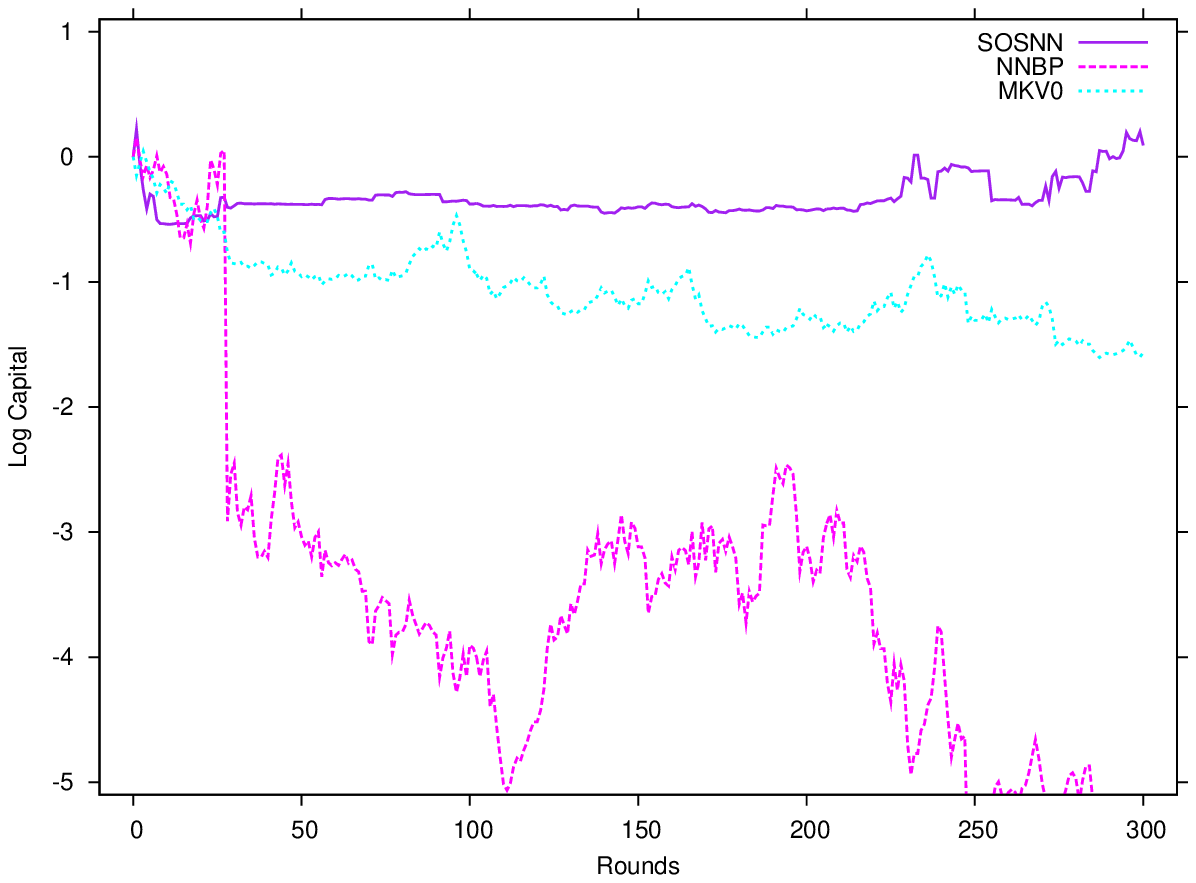}
\caption{NTT}
\label{fig:5}
\end{center}
\end{minipage}
\end{figure}

In Figure \ref{fig:2} we show the movements of closing prices of each company during the investing period. 
In Figures \ref{fig:3}-\ref{fig:5} we show the log capital processes of the results shown in Table \ref{real} to compare 
the performance of each strategy. Figure \ref{fig:3} is for SONY, Figure \ref{fig:4} is for Nomura Holdings and Figure \ref{fig:5} is for NTT. 
For SOSNN we plotted the result of $\lag $ and $M$ 
that gave the best performance at $n=300$ (the bottom row of the three rows) in Table \ref{real}.


As we see from above figures, NNBP which shows competitive performance 
for two linear models in Section \ref{sec:sim} gives the worst result. Thus it is obvious that 
the network has failed to capture trend in the betting period even if it fits in the training period. 
Also the results are favorable to SOSNN if we adopt appropriate numbers for $\lag$ and $M$. 

\section{Concluding remarks}
\label{sec:remarks}

We proposed investing strategies based on neural networks which directly consider 
Investor's capital process and are easy to implement in practical applications. We also 
presented numerical examples for simulated and actual stock price data to show advantages 
of our method.

In this paper we only adopted normalized values of past Market's movements for the input
$\bm{u}_{n-1}$ while  
we can use any data available before the start of round $n$ as a part of the input
as we mentioned in Section \ref{subsec:design-of-network}. 
Let us summarize other possibilities considered in existing researches on
financial prediction with neural networks. 
The simplest choice is to use raw data without any normalization as in \cite{hanias}, 
in which they analyze time series of Athens Stock index to predict future daily index. 
In \cite{bernd} they adopt price of FAZ-index 
(one of the German equivalents of the American Dow-Jones-Index), moving averages for 5, 10 and 90 days, 
bond market index, order index, US-Dollar and 10 successive FAZ-index prices as inputs 
to predict the weekly closing price of the FAZ-index. 
Also in \cite{khoa} they use 12 technical indicators 
to predict the S\verb|&|P 500 stock index one month in the future. From these  researches we see
that for longer prediction terms (such as monthly or yearly), 
longer moving averages or seasonal indexes become more effective. 
Thus those long term indicators may not have much effect in daily price prediction 
which we presented in this paper. 
On the other hand, adopting data which seems to have a strong correlation with closing prices of Tokyo Stock Exchange 
such as closing prices of New York Stock Exchange of the previous day 
may increase Investor's capital processes
presented in this paper. 
Since there are numerical difficulties in optimizing neural networks, 
it is better to use small number of effective inputs  for a good performance. 

Another important generalization of the method of this paper 
is to consider portfolio optimization. 
We can easily extend the method in this paper 
to the betting on  multiple assets. 
Let the output layer of the network have $P$ neurons as shown in Figure \ref{fig:6} 
and the output of each neuron is expressed as $y_{h}^{3}$, $h=1,\dots,P$. 
Then we obtain a vector $\bm{y}^{3}=(y_{1}^{3}, \dots, y_{P}^{3})$ of outputs.
The number of neurons $P$ refers to the number of different stocks Investor invests. 

\begin{figure}[thbp]
\vspace*{5mm}
\begin{center}
\includegraphics[width=75mm]{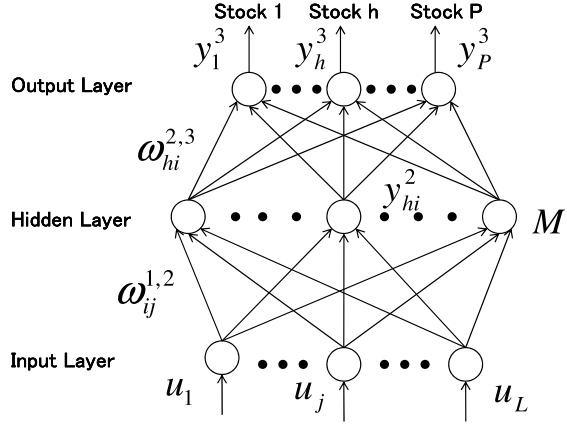}
\end{center}
\caption{Three-layered network for portfolio cases}
\label{fig:6}
\vspace*{5mm}
\end{figure}

Investor's capital after round $n$ is written as 
\[
\cK_{n}=\cK_{n-1}(1+\displaystyle\sum_{h=1}^{P}f_h(\bm{u}_{n-1}, \bm{\omega}_{h}) x_{n,h}),
\]
where 
\[
\bm{\omega}_{h}=
\big( (\omega_{ij}^{1,2})_{i=1,\dots,M,\,j=1,\dots,\lag}, (\omega^{2,3}_{hi})_{i=1,\dots,M}\big).
\]
Thus also in portfolio cases we see that our  method is easy to implement 
and we can evaluate Investor's capital process in practical applications. 

\section*{Appendix}
Here we discuss training error in the training period of NNBP. 
In this paper we set the threshold $\mu$ for ending the iteration 
to  $10^{-2}$, 
while the value 
commonly adopted in many previous researches is smaller, for instance, $\mu=10^{-4}$. 
We give some details on our choice of $\mu$.

\begin{figure}[htpb]
\begin{minipage}{0.5\hsize}
\begin{center}
\includegraphics[width=70mm]{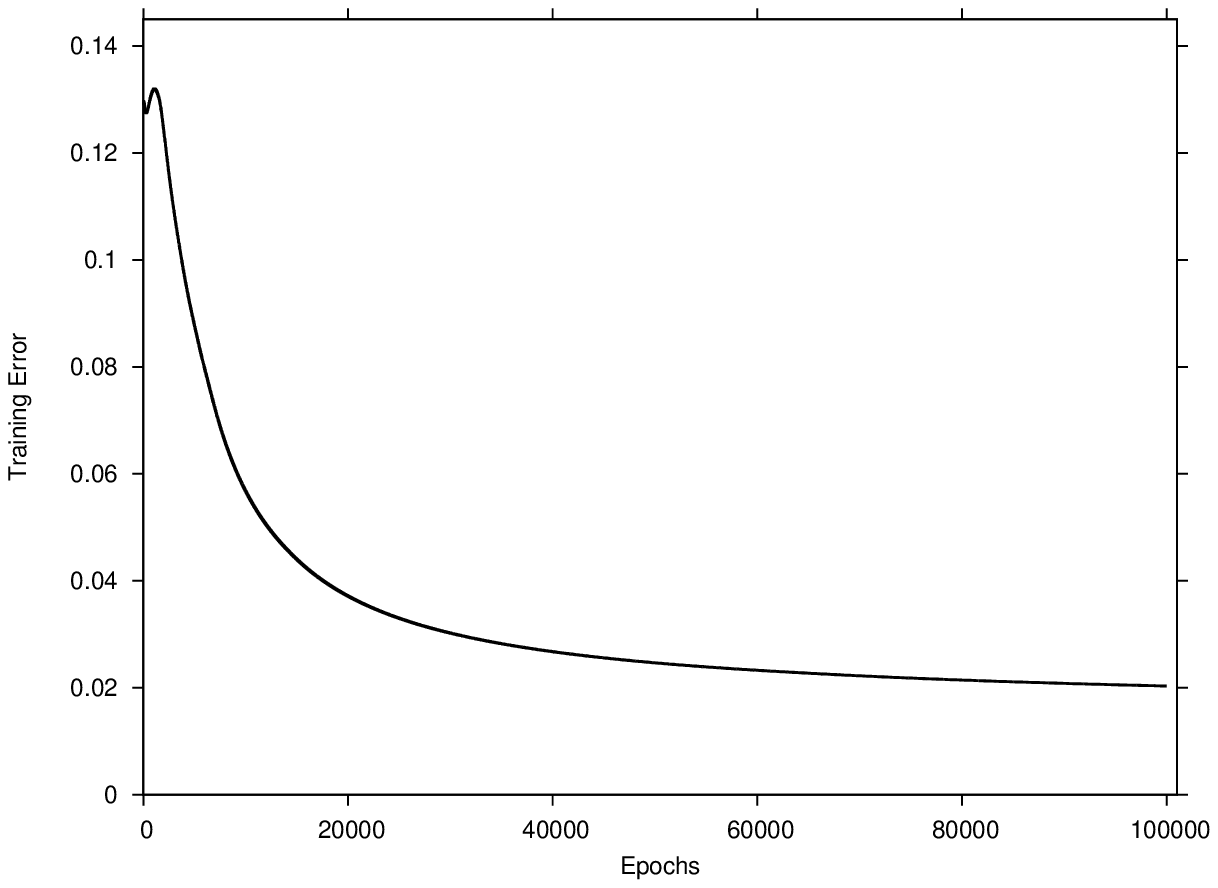}
\caption{training error (epochs)}
\label{fig:6}
\end{center}
\end{minipage}
\begin{minipage}{0.5\hsize}
\begin{center}
\includegraphics[width=70mm]{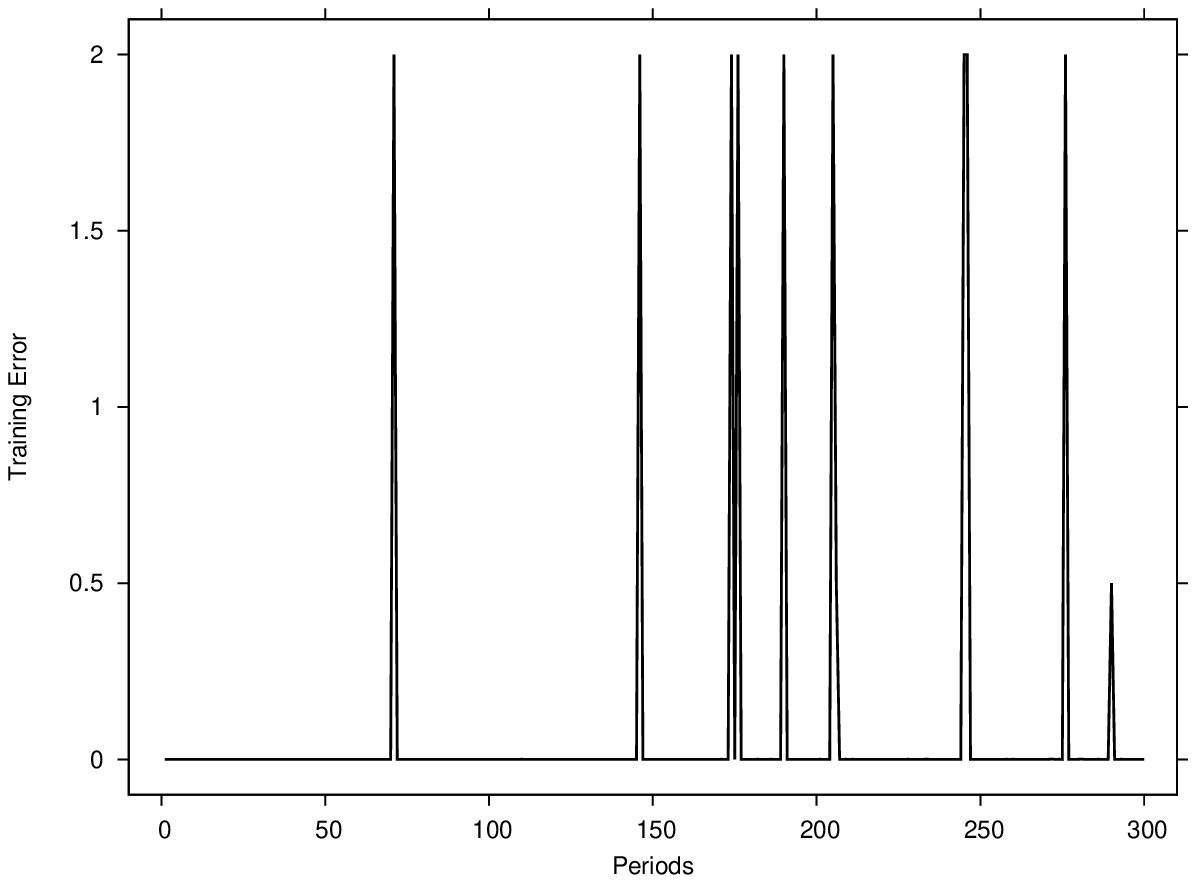}
\caption{training error (periods)}
\label{fig:7}
\end{center}
\end{minipage}
\end{figure}

Let us examine the case of Nomura Holdings in Section \ref{sec:real}. In Figure \ref{fig:6} we show 
the training error after each step of iteration in the training period calculated with (\ref{eq:te}). 
While the plotted curve has a typical shape as those of previous researches, 
it is unlikely that the training error becomes less than $10^{-2}$. 
Also in Figure \ref{fig:7} we plot $E_k=\frac{1}{2}(T_k-{}^{k}\!y^{3})^2$ for each $k$ 
calculated with parameter values $\bm{\omega}$ after learning. 
We observe that the network fails to fit for some points (actually $9$ days out of $300$ days) 
but perfectly fits for all other days. It can be interpreted that the network ignores some outliers 
and adjust to capture the trend of the whole data. 

{}

\end{document}